\documentclass{aastex631}

\usepackage{bm}
\usepackage{CJK}
\usepackage{pifont}
\usepackage{amsmath}
\usepackage{booktabs}
\usepackage{algorithm}
\usepackage{algpseudocode}
\usepackage{url}
\usepackage{enumitem}
\setlist[itemize]{noitemsep, topsep=0pt}

\definecolor{forestgreen}{HTML}{568203}
\newcommand{\cmark}{\textcolor{forestgreen}{\ding{51}}}
\newcommand{\xmark}{\textcolor{red}{\ding{55}}}

\def\C{{\mathbb{C}}}
\def\E{{\mathbb{E}}}
\def\R{{\mathbb{R}}}

\def\Dcal{{\mathcal{D}}}
\def\Fcal{{\mathcal{F}}}
\def\Ncal{{\mathcal{N}}}

\def\kbm{{\bm{k}}}

\def\nbm{{\bm{n}}}

\def\Abm{{\bm{A}}}

\def\Ibm{{\bm{I}}}

\def\Sbm{{\bm{S}}}

\def\Vbm{{\bm{V}}}

\def\E{{\mathbb{E}}}

\begin{document}

\title{POLISH'ing the Sky: Wide-Field and High-Dynamic Range Interferometric Image Reconstruction with Application to Strong Lens Discovery}

\correspondingauthor{Liam Connor}
\email{liam.connor@cfa.harvard.edu}

\author[0000-0002-7622-3548]{Zihui Wu}
\affiliation{Department of Computing and Mathematical Sciences, California Institute of Technology (Caltech), Pasadena, CA, USA}
\author[0000-0002-7587-6352]{Liam Connor}
\affiliation{Center for Astrophysics $\mid$ Harvard \& Smithsonian, Cambridge, MA, USA}
\author[0009-0008-5043-6220]{Samuel McCarty}
\affiliation{Center for Astrophysics $\mid$ Harvard \& Smithsonian, Cambridge, MA, USA}
\author[0000-0003-0077-4367]{Katherine L. Bouman}
\affiliation{Department of Computing and Mathematical Sciences, California Institute of Technology (Caltech), Pasadena, CA, USA}



\begin{abstract}

Radio interferometry enables high-resolution imaging of astronomical radio sources by synthesizing a large effective aperture from an array of antennas and solving a deconvolution problem to reconstruct the image.
Deep learning has emerged as a promising solution to the imaging problem, reducing computational costs and enabling super-resolution. 
However, existing DL-based methods often fall short of the requirements for real-world deployment due to limitations in handling high dynamic range, large field of view, and mismatches between training and test conditions.
In this work, we build upon and extend the POLISH framework, a recent DL model for radio interferometric imaging.
We introduce key improvements to enable robust reconstruction and super-resolution under real-world conditions: (1) a patch-wise training and stitching strategy for scaling to wide-field imaging and (2) a nonlinear arcsinh-based intensity transformation to manage high dynamic range.
We conduct comprehensive evaluations using the T-RECS simulation suite with realistic sky models and point spead functions (PSF), and demonstrate that our approach significantly improves reconstruction quality and robustness.
We test the model on realistic simulated strong gravitational lenses and show that lens systems with Einstein radii near the PSF scale can be recovered after deconvolution with our POLISH model, potentially yielding 10$\times$ more galaxy-galaxy lensing systems from the Deep Synoptic Array (DSA) survey than with image-plane CLEAN. 
Our results highlight the potential of DL models as practical, scalable tools for next-generation radio astronomy.

\end{abstract}



\section{Introduction} \label{sec:intro}

High-resolution imaging of astronomical radio sources is a fundamental problem in astronomy.
The resolution of a single-dish telescope is inherently limited by its aperture size (i.e., dish diameter), where the diffraction limit is given by roughly $\lambda/D$. Radio interferometers are used to achieve higher resolution by combining an array of antennas into a synthetic telescope \citep{ryle1946solar}.
An array of antennas together create a synthetic aperture that can be much larger than any single telescope dish, leading to a much higher spatial resolution.
Thanks to its ability to probe extremely small spatial scales, radio interferometry has unlocked numerous scientific discoveries, such as providing the first-ever direct image of a black hole \citep{event2019first_paper4}, evidence for dark matter subhalos in strong gravitational lenses \citep{mckean2025}, and the radio afterglow to gravitational wave events \citep{hallinan2017radio}.

The raw measurements in interferometry are given by coherently combining radio waves from all the antennas.
Each measurement between a pair of antennas samples a distinct spatial frequency in the Fourier domain.
In the image domain, recovering an image from interferometric measurements amounts to solving an image deconvolution problem where the convolution kernel is the point spread function (PSF) of the interferometer.
The PSF characterizes the response of the interferometer for point sources (i.e., impulse response) and is determined by the placement of the antennas and observing frequency.
One important aspect is the maximum ``baseline'' (i.e., the largest separation between two antennas) of the array, which is inversely proportional to the diffraction limit and thus the resolution of the interferometer.
Another important aspect is the number of antennas and distribution of unique baseline vectors.
A non-redundant array with more antennas leads to more combinations of interferometric measurements and thus a denser sampling pattern of the Fourier domain, leading to better image quality.

The upcoming Deep Synoptic Array (DSA)  will be the most powerful radio survey observatory to date \citep{hallinan2019dsa2000radiosurvey}. Its unprecedented sensitivity is predicted to increase the number of known radio sources by nearly two orders of magnitude in its first 5-year survey. 
With 1650 antennas, the DSA's PSF will be extremely well-behaved and thus more easily deconvolved.
However, the unparalleled image quality of DSA comes with significant computational challenges.
The raw data throughput of DSA is expected to be over 80 Tb/s, making it impossible to save data to disk for later analysis.
Traditional image reconstruction methods are limited in image quality (e.g., CLEAN \citep{hogbom1974aperture}) or efficiency to meet the expectation of DSA (e.g., optimization-based approaches).
Therefore, an efficient and robust image processing algorithm that quickly turns raw dirty observations into high-fidelity reconstructions is essential for the success of DSA.

\begin{table*}
\caption{\textbf{Existing works on deep learning for radio interferometry.} See App.~\ref{app:table_explanation} for detailed explanations.}
\label{TAB:COMPARISON}
\resizebox{\textwidth}{!}{
\begin{tabular}{cccccc}
\toprule
Methods & Max. image size & Max. dynamic range & Domain & Data & Model mismatch from calib. error \\ \midrule
POLISH \citep{connor2022deep} & $2,048$ & $\sim10^2$ & Image-plane & Synthetic (Gaussian) \& Real VLA & Synthetic \\
Radionets \citep{schmidt2022deep} & $1,024$ & $\sim5$ & Visibility-based & Synthetic (Gaussian sources) & -- \\
Radionets \citep{geyer2023deep} & $128$ & $\sim5$ & Visibility-based & Synthetic (Gaussian sources) & -- \\
Deflation Net \citep{chiche2023deep} & $256$ & $10^2$ & Image-plane & Synthetic (point sources) & -- \\
R2D2 \citep{aghabiglou2024r2d2} & $512$ & $5\cdot10^5$ & Image-plane & Synthetic (based on 4 real images) & -- \\
GU-Net \citep{mars2025learned} & $256$ & $\sim6\cdot10^2$ & Image-plane & Synthetic (IllustrisTNG simulations) & -- \\
RI-GAN \citep{mars2025generative} & $360$ & $\sim6\cdot10^2$ & Image-plane & Synthetic (Gaussian sources) & -- \\ \midrule
POLISH+ \& POLISH++ (Ours) & $12,960$ & $\sim10^6$ & Image-plane & Synthetic (Gaussian sources \& strong lenses) & Synthetic \\
\bottomrule
\end{tabular}}
\end{table*}

Recently, deep learning-based methods have been proposed for radio interferometric imaging \citep{connor2022deep, schmidt2022deep, geyer2023deep, chiche2023deep, mars2025learned, mars2025generative}.
These methods demonstrate a strong potential for efficient, accurate inversion of interferometric measurements and super-resolution beyond the inherent diffraction limit, making them promising candidates for satisfying the practical needs of DSA. Super-resolution is possible if one has a good prior on the PSF shape and basic knowledge of the observed sky (e.g. non-negative, sparse, and so on), both of which are true in radio interferometry.
However, these methods have not been tested to the level of real-world deployment and evaluation in several aspects.
Tab.~\ref{TAB:COMPARISON} provides a summary over four aspects of our interest in this work (see App.~\ref{app:table_explanation} for more details).
The existing methods are mostly tested on relatively small image sizes ($<1,000$ pixels per side) with low dynamic range ($<10^3$).
In contrast, real-world wide-field imaging on a sensitive survey telescope like DSA will produce images with more than $10,000$ pixels and a dynamic range on the order of $10^6$ \citep{hallinan2019dsa2000radiosurvey}. 
Moreover, existing works have used simple Gaussian-shaped galaxies, while  more complex shapes are expected in practice. For example, a rigid prior on the ``true sky'' will preclude the discovery of objects like strong gravitational lenses or galaxies with extended morphologies.
There is also expected to be a training/testing generalization gap due to real-world effects such as the ionosphere, pointing errors, gain variation, and other calibration errors that lead to differences in the assumed PSF and the true PSF during a given observation.
As shown in the last column of Tab.~\ref{TAB:COMPARISON}, most existing works do not consider any model mismatch between training and inference, so it remains unclear how the methods of these works will perform when such a generalization gap exists.

In this work, we provide both improved techniques and further evaluations towards real-world deployment and evaluation of deep learning-based methods for radio interferometric imaging.
Specifically, we tackle two prominent problems in the deployment of DL methods to real-world interferometry imaging problems with large FOV like DSA: (1) the prohibitively high dimensionality of the target image and (2) the prohibitively high dynamic range due to a large number of radio sources.
We use the POLISH as our base architecture \citep{connor2022deep}.
For (1), we employ a patch-wise training and inference technique that allows us to train on smaller patches and stitch the patch-wise predictions into a larger reconstruction.
For (2), we leverage a simple nonlinear transform based on inverse hyperbolic functions that can reduce the dynamic range for training the neural network.
To comprehensively evaluate these techniques (and DL models in general), we use large-dimension images with extremely high dynamic ranges based on the T-RECS simulation.
Our evaluation is based on not only image quality but also the ability to detect radio sources, as well as estimating their shapes and fluxes, motivated by the real-world needs for radio interferometry.

We further demonstrate the model's ability to reconstruct and super-resolve realistic strongly lensed objects.
Strong gravitational lensing has important applications in astrophysics and cosmology, including constraining $H_0$ via time-delay studies, studying dark matter and dark matter substructure at cosmological distances, and magnifying extremely distant and faint sources \citep{Treu2010, Vegetti2024}.
A main challenge in strong lens science is the identification of lens systems, which are buried in huge amounts of wide-field survey data.
One of the most important parameters for determining a survey's ability to find lenses is the resolution, as generally the multiple images produced by strong lensing need to be resolved for reliable identification.
The upcoming DSA survey is expected to discover roughly $10^4-10^5$ strong lenses, up to a three-order-of-magnitude increase in the radio lens sample, which will be mostly galaxy group and cluster scale systems due to the relatively poor resolution of the DSA \citep{mccarty2024strong}.
However, the strong lens yield of the DSA can be increased significantly (nearly an order of magnitude) if lenses with separations below the PSF scale can be recovered \citep{mccarty2024strong}.
We wish to test if \texttt{POLISH} enabled super-resolution, when combined with a lens finder, can reliable recover these lenses. 

We also test the robustness and flexibility of the proposed method by training on data generated with an idealized PSF and evaluating on data corrupted by perturbed PSFs that fall outside the training distribution. This setup probes the model’s sensitivity to realistic calibration and modeling errors, such as those arising from ionospheric phase distortions or antenna mispointing. In addition to demonstrating robustness to such mismatches, we show that POLISH++ can be efficiently adapted to new PSF distributions via fine-tuning, enabling rapid deployment across different pointings and observing conditions without retraining from scratch.

The manuscript is organized as follows.
We first review the relevant background in Sec.~\ref{sec:background}, including the problem formulation of radio interferometric imaging and existing approaches.
We then outline our method in Sec.~\ref{sec:method}, focusing mainly on the additional techniques that we build upon POLISH.
We finally present a comprehensive evaluation of our approach in Sec.~\ref{sec:experiments} on (1) images generated with a realistic sky model, (2) images with strongly lensed objects, and (3) images convolved with randomly perturbed PSFs.
\section{Background} \label{sec:background}

\subsection{Interferometric Imaging as Deconvolution}

In radio interferometry, the goal is to reconstruct an image of the sky from the interferometric measurements, which are often referred to as ``visibilities'' and denoted by $\Vbm \in \C^m$. These are complex time-averaged correlations between the voltage signals at 
antennas $i$ and $j$, i.e. $V_{i,j} = \langle v_i v^*_j\rangle$.
The relationship between the measurements and the underlying target image $\Ibm_\text{sky} \in \R^n$ can be formulated as
\begin{align*}
    \Vbm = \Sbm \Fcal(\Ibm_\text{true} + \nbm)
\end{align*}
where $\Sbm \in \{0, 1\}^{m \times n}$ is a binary sampling matrix, $\Fcal(\cdot)$ denotes the 2D Fourier transform, and $\nbm$ represents the background noise of the sky, usually modeled as independent and identically distributed (i.i.d.) Gaussian noise.
Taking the inverse Fourier transform of both sides and using the Fourier convolution theorem, we have that
\begin{align*}
    \Ibm_\text{dirty} = \Fcal^{-1}(\Sbm) * (\Ibm_\text{true} + \nbm) = \kbm * (\Ibm_\text{true} + \nbm)
\end{align*}
where $\Ibm_\text{dirty} := \Fcal^{-1}(\Vbm)$ is the dirty sky image given by naively inverting the observed visibilities, $*$ denotes 2D spatial convolution, and $\kbm := \Fcal^{-1}(\Sbm)$ represents the PSF (i.e., blur kernel) of the interferometer.
For achieving super-resolution, it is common to further assume that the actual dirty sky image is a lower-resolution version of the underlying target, i.e.,
\begin{align}
\label{eqn:forward_model}
    \Ibm_\text{dirty} = \left[ \kbm * (\Ibm_\text{true} + \nbm) \right] \downarrow_s
\end{align}
where $\downarrow_s$ represents the down-sampling operation by a factor of $s$.
Interferometric imaging amounts to designing an inverse mapping that predicts $\Ibm_\text{true}$ with input $\Ibm_\text{dirty}$.

\subsection{Existing Techniques for Radio Interferometric Imaging}

\noindent \textbf{CLEAN} \quad CLEAN has been the standard reconstruction method in radio astronomy for decades \citep{hogbom1974aperture, clark1980efficient}.
By assuming a sky model made of point sources, CLEAN iteratively removes the brightest sources in the dirty image $\Ibm_\text{dirty}$ by subtracting the PSF $\kbm$ at the location of that source, and adds a point source to the sky model at the corresponding location.
The added point sources are scaled by the fluxes of the corresponding sources in the dirty image.
At the final step, the sky model with all added point sources is convolved with the a Gaussian blur kernel (i.e., a restoring beam) to provide a smooth reconstruction.
Despite its wide application and intuitive mechanism, CLEAN has two limitations that fall short of the need in more advanced interferometric imaging.
First, the artifacts in the dirty image may not have the same shape as the PSF, such as the artifacts around diffuse or spiral sources.
Therefore, a simple subtraction of PSF often cannot fully remove the artifacts of sources with complex shapes.
Second, due to the final blurring step, the resolution of the CLEAN reconstruction is limited by the angular scale of the central component of the PSF.
It is thus impossible to achieve super-resolution beyond the resolution of the PSF, if the standard restoring beam procedure is applied.

\noindent \textbf{Regularized Maximum Likelihood (RML)} \quad RML is an optimization framework based on the Bayesian probabilistic formulation.
The reconstruction problem can be formulated as
\begin{align}
\label{eqn:rml}
    \Ibm^\ast = \arg\min_{\Ibm}\frac{1}{2}\|\Abm(\Ibm)-\Ibm_\text{dirty}\|_2^2 + \gamma \cdot r(\Ibm),
\end{align}
where $\Abm(\Ibm) := \left[ \kbm * (\Ibm + \nbm) \right] \downarrow_s$ denotes the forward model, $r(\Ibm): \R^n \to \R_+$ is a regularization term, and $\gamma > 0$ controls the regularization strength.
Common choices of regularization include the $l_1$ norm, maximum entropy, total variation \citep{rudin1992nonlinear}, and so on.
RML has found success in many radio astronomy applications, including its use as part of the imaging pipeline for the first-ever image of a black hole \citep{event2019first_paper4}.
It has also been shown that RML has super-resolution ability \citep{abdullah2019wideband}.
However, as solving (\ref{eqn:rml}) often requires a computationally expensive iterative procedure, it is not suitable for high-throughput application scenarios like DSA.

\noindent \textbf{Deep Learning Approaches} \quad Over the past decade, deep learning (DL) methods have become state-of-the-art for image deconvolution and super-resolution \citep{xu2014deep, yan2016blind, dong2016image, wang2021deep, zhang2022deep}.
More recently, DL-based methods have also gained popularity for radio interferometric imaging due to their ability to perform fast, high-fidelity reconstructions from incomplete or noisy visibility data.
The existing methods can be categorized into three main classes.
The first class consists of end-to-end models that directly map dirty images to clean reconstructions using deterministic feed-forward neural networks. 
Examples include POLISH \citep{connor2022deep}, Radionets \citep{schmidt2022deep, geyer2023deep}, Deflation Net \citep{chiche2023deep}, and GU-Net \citep{mars2025learned}, which are trained to invert the imaging process without explicitly modeling the measurement operator during inference.
The second class leverages pre-trained deep denoisers as priors within iterative optimization frameworks.
For instance, BC-RED \citep{sun2019block} incorporates a CNN denoiser into the Regularization by Denoising (RED) framework \citep{romano2017little} via block-coordinate updates. 
Similarly, DeepShape \citep{tripathi2025deepshape} and AIRI \citep{terris2022image, dabbech2022first} employs the deep denoisers within the Half-Quadratic Splitting (HQS) and Forward-Backward (FB) algorithms, respectively, leveraging the plug-and-play priors methodology \citep{venkatakrishnan2013plugandplay}.
Unrolling networks that combine iterative updates and deep learning, such as R2D2 \citep{aghabiglou2024r2d2}, have also been explored.
The third class comprises deep generative models, which aim to characterize the full posterior distribution of plausible reconstructions.
This includes both adversarially trained models such as RI-GAN \citep{mars2025generative}, which couples a generator with a data-consistency module, and score-based models that enable sampling-based posterior inference \citep{remy2023probabilistic}.
\section{Method} \label{sec:method}

\subsection{Preliminaries: POLISH}

POLISH is a deep learning model for radio interferometry proposed by \cite{connor2022deep}.
In POLISH, an end-to-end convolutional neural network (CNN), which we denote as $G_\theta(\cdot)$, is trained to learn a direct mapping from the low-resolution dirty image (given as input) to the high-resolution target image with a training dataset of dirty and true image pairs $\Dcal_\text{train}:=\left\{\left(\Ibm_\text{dirty}^{(i)}, \Ibm_\text{true}^{(i)}\right)\right\}_{i=1}^{N}$ and the $\ell_1$ loss defined as
\begin{align*}
    \theta^\ast = \arg\min_{\theta} \frac{1}{N} \sum_{i=1}^{N} \left\|G_\theta\left(\Ibm_\text{dirty}^{(i)}\right) - \Ibm_\text{true}^{(i)}\right\|_1.
\end{align*}
The architecture of POLISH is modified based on WDSR \citep{fan2018wide, yu2018wide}.
As shown in \cite{connor2022deep}, POLISH exhibits several advantages over classical radio interferometric imaging methods.
First, as a fully convolutional and single-pass model, it offers high computational efficiency, making it well-suited for the quasi-real-time data rates expected from next-generation interferometers such as DSA.
Second, it achieves super-resolution beyond the PSF resolution, allowing it to recover fine-scale structures in astronomical sources that are typically unresolved by CLEAN-based methods. 
Third, the model produces visually cleaner reconstructions with significantly fewer sidelobe artifacts, enabling more accurate galaxy detection and parameter estimation.
Finally, because it is trained end-to-end on realistic simulations, POLISH can implicitly learn complex mappings that would be difficult to capture with traditional iterative algorithms.

\begin{figure}[t]
\centering
\includegraphics[width=\linewidth]{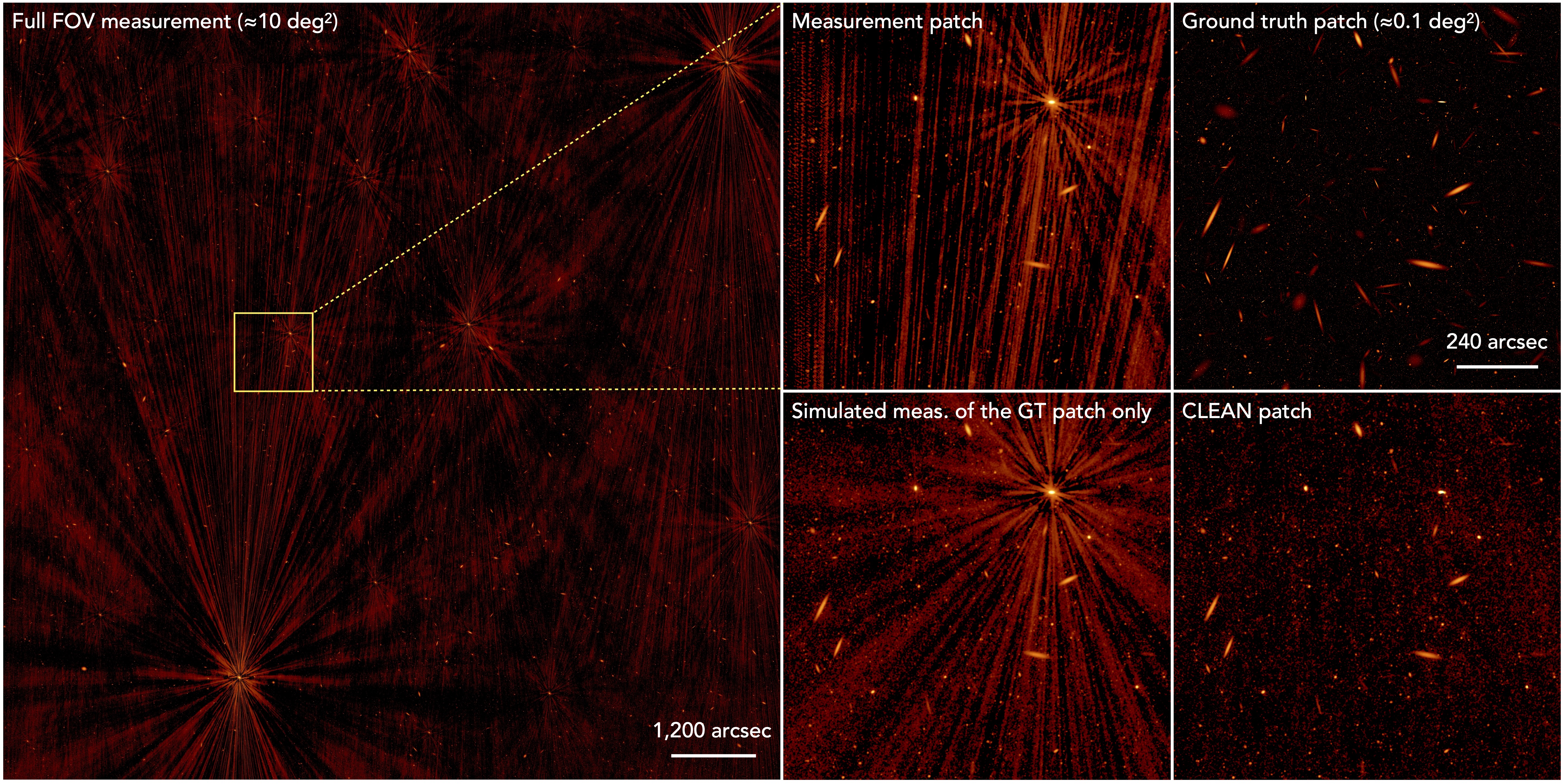}
\caption{
\textbf{Illustration of patch-wise measurements and cross-patch contamination in wide-field interferometric imaging.}
Left: the full–field-of-view dirty measurement  ($12,960\times12,960$ pixels), with the highlighted region indicating the patch under consideration ($1,296\times1,296$ pixels). Top right: the ground-truth sky patch. Top middle: the corresponding patch extracted from the full-FOV dirty measurement. Bottom middle: a simulated measurement obtained by convolving the ground-truth patch alone with the PSF. Bottom right: the corresponding patch of the CLEAN reconstruction from the full-FOV measurement. The measurement patch extracted from the full FOV contains strong artifacts induced by bright sources outside the patch, visible as extended streaks from PSF sidelobes (e.g., from sources below the patch). In contrast, these nonlocal artifacts are absent when simulating the measurement from the ground-truth patch alone, highlighting  that the standard forward model does not apply locally at the patch level and motivating learning-based approaches that can handle such cross-patch contamination.
}
\label{fig:patch_lr}
\end{figure}

\subsection{Patch-Wise Processing}

A direct bottleneck caused by the large FOV of DSA is the large dimensionality of the underlying image.
For a FOV of roughly 10 square degrees and an angular resolution of order 1$''$, the resulting image can exceed $10{,}000 \times 10{,}000$ pixels.
With an image size exceeding $10^8$ pixels, a single float32 input already requires approximately 400 MB of GPU memory.
During training, intermediate feature maps dominate memory usage; for example, storing activations with only 32 channels at full resolution would require over 12 GB for a single layer, which is further multiplied during backpropagation due to gradient storage.
Consequently, the total memory footprint can easily exceed the capacity of a single modern GPU (typically 20–80 GB) even with batch size 1.
This may necessitate either model parallelism (i.e., distributing a model across multiple GPUs) or a lower training precision, leading to extra complexities and potentially worse performance.
Moreover, obtaining a well-trained network usually require thousands or more training samples.
The large dimension of each training sample may also introduces bottlenecks in disk space and input/output (I/O), leading to limited training efficiency.

To tackle these challenges, we propose to train and infer on significantly smaller image patches instead.
For each pair of dirty sky image (measurement) and clean sky image (ground truth) in the training set, we dissect the images into $J$ non-overlapping patches with matching locations. 
After doing so, one training pair $\left(\Ibm_\text{dirty}^{(i)}, \Ibm_\text{true}^{(i)}\right)$ becomes a set of $J$ training pairs of smaller size $\left\{\left(\Ibm_\text{dirty}^{(i),[j]}, \Ibm_\text{true}^{(i),[j]}\right)\right\}_{j=1}^{J}$.
The total training set is the union of all patches of all training images, i.e., $\Dcal_\text{train}:=\bigcup_{i=1}^{N}\left\{\left(\Ibm_\text{dirty}^{(i),[j]}, \Ibm_\text{true}^{(i),[j]}\right)\right\}_{j=1}^{J}$.

Although this approach may appear to be straightforward, we note that it is substantially different from directly training on smaller images, as was done in \cite{connor2022deep}.
We provide some visual examples in Fig. \ref{fig:patch_lr} to highlight this difference.
For an example test image, we visualize one of its patches (top right), the corresponding patch of the full-FOV dirty sky measurement (top middle), and the simulated measurement given by convolving the PSF with the ground truth patch only (bottom middle).
Note that the measurement patch contains significant artifacts caused by bright sources from nearby patches.
In this case, there should be bright sources from patches below this patch as shown by the dark streaks coming from the bottom caused by the PSF side lobes.
However, such artifacts do not exist for the simulated measurement of the ground truth patch only (top bottom).
From a forward model perspective, although (\ref{eqn:forward_model}) governs the relationship between $\Ibm_\text{dirty}$ and $\Ibm_\text{true}$ over the entire FOV, no explicit relationship exists between $\Ibm_\text{dirty}^{[j]}$ and $\Ibm_\text{true}^{[j]}$.
One important evaluation of our interest is whether deep learning-based methods still work when being trained on the patch level.
This is particularly important for wide-field surveys that DSA will conduct because the dimension of the dirty sky images can easily become intractable for training on the entire FOV.

\subsection{Arcsinh-Based Transformation}

\begin{figure}[t]
\centering
\includegraphics[width=\linewidth]{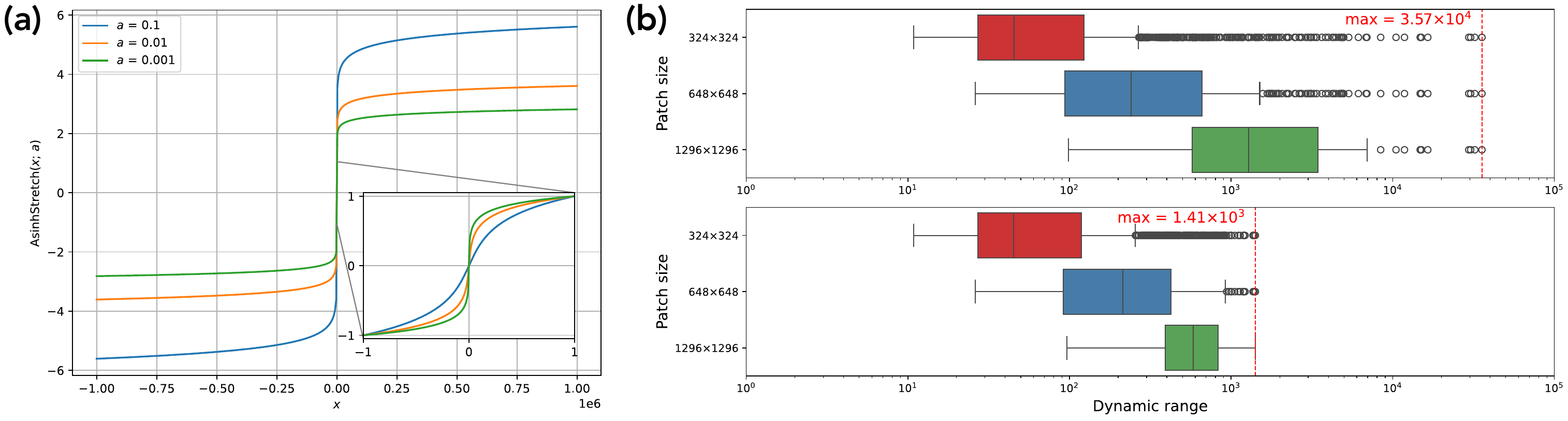}
\caption{
\textbf{Visualization of the nonlinear transform used for dynamic range reduction.}
\textbf{(a)} $\mathsf{AsinhStretch}(x; a)$ defined in (\ref{eqn:asinh}) with $a \in \{0.1, 0.01, 0.001\}$.
\textbf{(b)} Visualize of the dynamic range distribution of one test image before (top) and after (bottom) applying $\mathsf{AsinhStretch}(x; a=0.001)$ .
One can see that the maximum dynamic range is reduced by over one order of magnitude after applying the transformation. 
}
\label{fig:asinh}
\end{figure}

Another challenge due to the large the field-of-view (FOV) of DSA is that the dirty image will likely to have extremely high dynamic range (i.e., the ratio between the brightest and dimmest pixels), which can reach around $10^5 \sim 10^6$ \citep{mars2025learned, mars2025generative}.
To overcome such a high dynamic range, we propose to utilize the following nonlinear transformation:
\begin{align}
\label{eqn:asinh}
    \mathsf{AsinhStretch}(x; a) := \frac{\operatorname{arcsinh}(x/a)}{\operatorname{arcsinh}(1/a)}
\end{align}
where $\operatorname{arcsinh}(x) = \ln\left(x + \sqrt{1+x^2}\right)$ is the inverse hyperbolic sine function.
In Fig.~\ref{fig:asinh} (a), we plot this transform with three values of $a$ with a zoom-in on $x \in [0, 1]$.
This transformation has a few desirable properties. 
First, due to its logarithmic form, it can compress the pixels on multiple orders of magnitude to the same order of magnitude, significantly reducing the dynamic range, as shown in Fig.~\ref{fig:asinh} (b).
Second, unlike gamma encoding in photography, it can handle both positive and negative values, making it suitable to process the dirty image $\Ibm_\text{dirty}$ which may contain negative values.

We propose to train in the transformed space given by this transformation.
The training loss is
\begin{align}
\label{eqn:loss}
    \theta^\ast = \arg\min_{\theta} \frac{1}{NJ} \sum_{i=1}^{N} \sum_{j=1}^{J} \left\|G_\theta\left(\mathsf{AsinhStretch}\left(\Ibm_\text{dirty}^{(i),[j]}; a_\text{dirty}\right)\right) - \mathsf{AsinhStretch}\left(\Ibm_\text{true}^{(i),[j]}; a_\text{true}\right)\right\|_1.
\end{align}
where $a_\text{dirty}$ and $a_\text{true}$ are hyperparameters that we set before training.
Once we have a trained model $G_{\theta^\ast}$, we use it for inference 
\begin{align*}
    \widehat{\Ibm}_\text{true} = \bigoplus_{j=1}^{J}\mathsf{AsinhStretch}^{-1}\left(G_{\theta^\ast}\left(\mathsf{AsinhStretch}\left(\Ibm_\text{dirty}^{[j]}; a_\text{dirty}\right)\right); a_\text{true}\right).
\end{align*}
where $\bigoplus_{j=1}^{J}$ is an operator that combines all $J$ patches back to a full FOV image.
The inverse transform $\mathsf{AsinhStretch}^{-1}(x; a) := a \cdot \sinh(x \cdot \operatorname{arcsinh}(1/a))$ is the inverse mapping of $\mathsf{AsinhStretch}(x; a)$, i.e., $\mathsf{AsinhStretch}^{-1}(\mathsf{AsinhStretch}(x; a); a) = x$.
\section{Experiments} \label{sec:experiments}

\begin{table*}
\caption{\textbf{A summary of methods for comparisons in the Experiments section (Sec.~\ref{sec:experiments}).}}
\label{tab:baselines}
\resizebox{0.8\textwidth}{!}{
\begin{tabular}{cccc}
\toprule
Methods & Learning-based & Patch-level training & Nonlinear transform \\ \midrule
CLEAN \citep{hogbom1974aperture} & \xmark & -- & \xmark \\
POLISH \citep{connor2022deep} & \cmark & \xmark & \xmark \\ \midrule
POLISH+ (this work) & \cmark & \cmark & \xmark \\
POLISH++ (this work) & \cmark & \cmark & \cmark \\
\bottomrule
\end{tabular}}
\end{table*}

\subsection{Experimental Setup}
\label{sec:experimental_setup}

We use the Tiered Radio Extragalactic Continuum Simulation (T-RECS) to produce ``ground truth'' images of the radio sky at DSA frequencies \citep{TRECSI,TRECSII}. T-RECS is a semi-analytic model of star-forming radio galaxies (SFRG) and radio Active Galactic Nuclei (AGN) that accurately reproduces observations. 
Catalogs of AGN and SFRGs from the T-RECS catalogs are used to populate 12,960$\times$12,960 pixel images with a pixel scale of 0.8784$''$, leading to an FoV of roughly 10 deg$^2$.
AGN are single-pixel sources in the ground truth image, as they are true point sources at this pixel scale. The SFRGs have elliptical Sérsic brightness profiles in T-RECS with an empirical size distribution \citep{TRECSI}.
An example ground truth patch is shown in Fig.~\ref{fig:patch_lr} (top right).
We then add Gaussian noise with mean zero and standard deviation 1\,$\mu$Jy to the ground truth images, which is the expected continuum sensitivity for a 1-hour integration with the DSA \citep{hallinan2019dsa2000radiosurvey}.
The dimmest detectable sources in our simulated images are thus roughly 5\,$\mu$Jy.
The maximum pixel values of our images are between 0.1 and 3\,Jy, giving a dynamic range between $10^4$ and $10^6$.
Next, we simulate the interferometric measurement process.
For this we use image-plane convolution with a physically realistic DSA PSF.
A major forward modeling effort is underway within the DSA collaboration, simulating end-to-end observations of the radio sky including different systematics (the ionosphere, calibration errors, primary beam effects, etc.).
We convolve the noisy ground truth image with the instrument PSF; this produces our measurement or ``dirty image''. The dirty images are downsampled by $2\times$ from the high-resolution ground truth images (i.e., $s=2$ in (\ref{eqn:forward_model})).

Bright sources limit the achievable dynamic range of radio interferometric imaging, and can be particularly damaging in the absence of ungridded visibility data. 
The DSA's real-time radio camera pipeline will mitigate this issue by ``peeling'' the brightest sources in each field.
Peeling refers to solving direction-dependent calibration parameters for individual sources and subtracting them from the raw visibility data before imaging \citep{Noordam, Mitchell_2008}.
This effectively reduces the image dynamic range and makes interferometric reconstructive less susceptible to calibration errors or pixelization effects.
While the exact threshold of peeling on the DSA will depend on available compute resources in the radio camera pipeline, it is likely that sources brighter than $\sim$\,100\,mJy will not be in the dirty images.
The main POLISH++ results presented here, which assume no peeling, are thus a worst-case-scenario in terms of dynamic range. 

With the dirty images in hand, we train POLISH models to map between the low-resolution dirty sky and the high-resolution ground truth sky.
In total, we use 18 images for training and 5 images for testing.
While this is a small number of images, the training set actually contains 28,800 non-overlapping images of 324$\times$324 pixels, due to our patch-wise processing approach.
There are a total of 6 million detectable galaxy examples in the training set from which POLISH can learn deconvolution.
We train three different POLISH models, progressively including the improvements described in Section \ref{sec:method}: a baseline POLISH, POLISH+, and POLISH++.
The differences are summarized in Table \ref{tab:baselines}.
We use the Common Astronomy Software Application (CASA) to create CLEAN images as our baseline for comparison \citep{TheCASATeam_2022}.
Because our simulated images are entirely in the image plane, we use CASA's \texttt{deconvolve} task, which is the minor cycle of the \texttt{tclean} task.
We use the `hogbom' deconvolver and a loop gain of 0.1. 
For all of the images, the CLEAN algorithm diverges at the $\sim$mJy level, corresponding to a typical CLEAN dynamic range of several hundred.
The CLEAN'ed images are then upsampled with interpolation by $2\times$ for comparison with the POLISH images. 

More implementation details can be found in App.~\ref{app:implementation_details}.

\subsection{Main Results}

\begin{figure*}[tb]
\centering
\includegraphics[width=\textwidth]{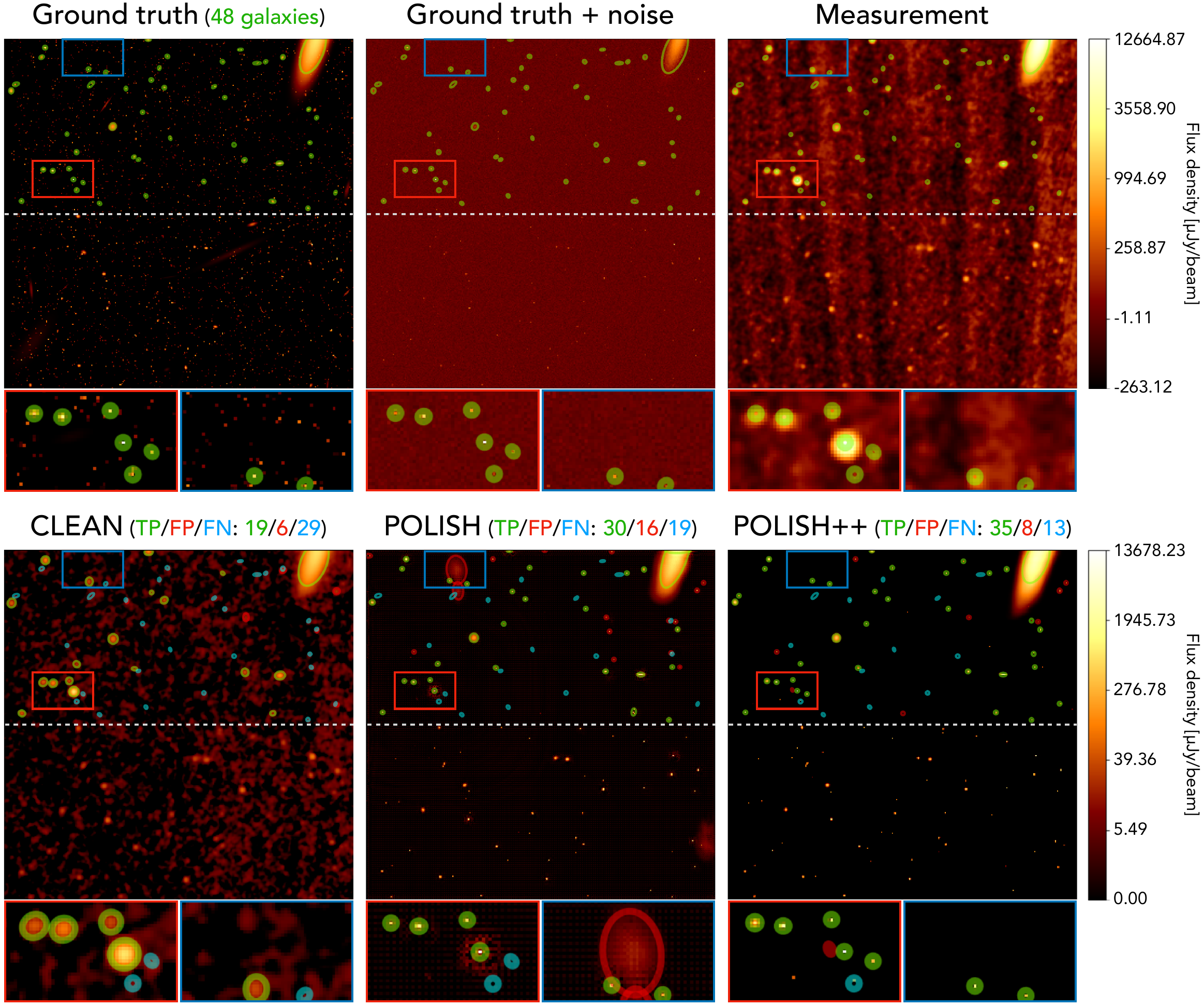}
\caption{
\textbf{Visual comparison of galaxy detection results across different reconstruction methods.}
We intentionally show detections only in the top half of each image to facilitate comparison of both the detections and the underlying image quality.
\underline{Top row:} ground truth sky (48 galaxies), ground truth with added Gaussian noise, and the corresponding low-resolution measurement.
\underline{Bottom row:} reconstructions obtained with CLEAN, POLISH, and POLISH++.
Detected sources are marked by ellipses, with true positives shown in green, false positives in red, and false negatives in blue (reported as TP/FP/FN in the caption of each method).
For each detection, the major and minor axis lengths of the ellipse are set to twice the corresponding FWHM values.
The cyan zoom-in regions highlight the improved spatial resolution achieved by POLISH+ and POLISH++ relative to the blurred CLEAN reconstructions, demonstrating their super-resolution capability.
The green zoom-in region illustrates background artifacts introduced by POLISH, whereas POLISH++ yields a markedly cleaner, artifact-free reconstruction with reduced hallucinations.
The top colorbar corresponds the pixel intensity scale of the ``Ground truth + noise'' and the ``Measurement'' panels, while the bottom colorbar corresponds to that of the other four panels.
}
\label{fig:visual_examples_detections}
\end{figure*}

\begin{figure}[t]
\centering
\includegraphics[width=\linewidth]{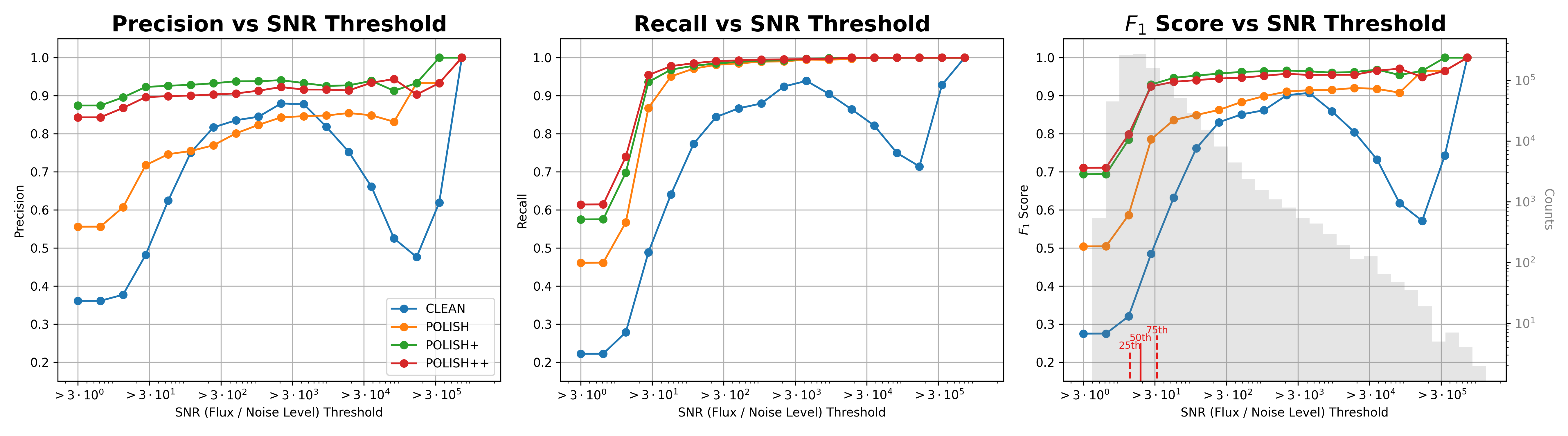}
\caption{
\textbf{Detection performance versus progressively larger signal-to-noise ratio (SNR) threshold.}
From left to right, the panels show precision, recall, and $F_1$ score for CLEAN, POLISH, POLISH+, and POLISH++ as the SNR threshold increases from 3 to 300,000.
For each subplot, the leftmost point roughly matches the corresponding values in Tab.~\ref{tab:baselines} as almost all sources have flux $>3$.
The background histogram in the rightmost panel (gray, logarithmic scale) visualizes the SNR distribution of all sources, indicating that most sources have relatively low SNRs.
Overall, POLISH+ and POLISH++ substantially outperform the baselines.
Furthermore, POLISH++ achieves a generally better tradeoff between precision and recall, leading to an improved $F_1$ score in the low-SNR-threshold regime which includes the majority of sources. 
}
\label{fig:accuracy_vs_snr}
\end{figure}

\begin{table}[t]
\centering
\caption{
\textbf{Quantitative results for source detection and shape \& flux estimation.}
}
\label{tab:main}
\resizebox{\linewidth}{!}{
\begin{tabular}{lcccccc}
\toprule
 & \multicolumn{3}{c}{Detection Accuracy} & \multicolumn{3}{c}{Shape \& Flux Estimation Accuracy (RMSE) on True Positive Detections} \\
 \cmidrule(lr){2-4} \cmidrule(lr){5-7}
 & Precision $\uparrow$ & Recall $\uparrow$ & $F_1$ score $\uparrow$ & Major Axis FWHM ($''$) $\downarrow$ & Minor Axis FWHM ($''$) $\downarrow$ & Flux (Jy/pixel) $\downarrow$ \\
\midrule
CLEAN & 0.3612 & 0.2220 & 0.2750 & 1.0046 & 0.7862 & $\mathbf{3.2625 \times 10^{-4}}$ \\
POLISH & 0.5560 & 0.4612 & 0.5042 & 0.9642 & 0.3219 & $1.9504 \times 10^{-3}$ \\
\midrule
POLISH+ & \textbf{0.8744} & 0.5751 & 0.6938 & \textbf{0.4335} & \textbf{0.1889} & $3.8411 \times 10^{-3}$ \\
POLISH++ & 0.8433 & \textbf{0.6142} & \textbf{0.7107} & 0.4654 & 0.2056 & $3.1703 \times 10^{-3}$ \\
\bottomrule
\end{tabular}
}
\end{table}

We present two sets of experiments that compare POLISH++ with CLEAN \citep{hogbom1974aperture} and POLISH \citep{connor2022deep} from a task-relevant perspective.
The first set focuses on the accuracy for detecting galaxies
(source detection), while the second set investigates the accuracy for estimating the shapes and fluxes of the detected galaxies (source parameter estimation).
The full quantitative results are provided in Tab. \ref{tab:main}.

\noindent \textbf{Source Detection} \quad 
We first assess the accuracy of detecting galactic sources using the \texttt{extract} function from the \texttt{SEP} package\footnote{\url{https://github.com/sep-developers/sep} (Lesser GNU Public License)}.
We use three standard metrics: precision, recall, and $F_1$ score, defined as
\begin{align*}
    \text{Precision}=\frac{\text{TP}}{\text{TP}+\text{FP}}, \qquad \text{Recall}=\frac{\text{TP}}{\text{TP}+\text{FN}}, \qquad F_1\text{ score}=\frac{2\cdot\text{TP}}{2\cdot\text{TP}+\text{FP}+\text{FN}},
\end{align*}
where TP, FP, FN denote the number of true positives, false positives, and false negatives, respectively.
A detection is considered successful if, for a certain galaxy in the noisy ground-truth image, there exists a detected galaxy in the reconstruction whose center lies within $N_\text{threshold}$ pixels of the ground-truth center.
To account for variations in galaxy size, we allow $N_\text{threshold}$ to depend on the size of the target galaxy rather than fixing it to a constant value.
The specific definition of $N_\text{threshold}$ is provided in App.~\ref{app:implementation_details}.
Given this criterion, a true positive is defined as the number of galaxies in the noisy ground truth that have successfully matched detections in the reconstruction.
A false negative is defined as the number of galaxies in the noisy ground truth that \emph{do not} have successfully matched detections in the reconstruction.
A false positive is defined as the number of galaxies in the reconstruction that \emph{are not} matched to any galaxy in the noisy ground truth.
We note that the ground truth detections are found by applying the same procedure on the noisy high-resolution image (i.e., $\Ibm_\text{true} + \nbm$ in (\ref{eqn:forward_model})) so the galaxies below the noise floor are not detected.

Overall, the results presented in Tab.~\ref{tab:main} demonstrate that the proposed POLISH+ and POLISH++ methods substantially outperform both the classical CLEAN baseline and the previous POLISH method.
Specifically, POLISH+ achieves significantly higher accuracies across all three metrics, validating the effectiveness of our patch-level learning approach.
By incorporating an additional nonlinear transformation, POLISH++ further improves upon POLISH+ in terms of $F_1$ score by achieving a better balance between precision and recall.
Note that the gain in recall is around 4\%, indicating that POLISH++ successfully recovers a significantly larger portion of the true galaxy population.

Furthermore, we investigate the dependence of detection performance on SNR.
Fig.~\ref{fig:accuracy_vs_snr} illustrates the precision, recall, and $F_1$ score metrics as the SNR threshold increases from 3 to 300,000.
Our proposed methods exhibit a significant performance improvement over the baselines at low SNR values.
Although the performance gap narrows as SNR increases, it is crucial to note that the low-SNR regime constitutes the vast majority of real-world cases, as indicated by the SNR histogram in the background of the rightmost panel. 

We also present some visual examples in Fig.~\ref{fig:visual_examples_detections}.
For the reconstructions in the second row, true positives, false positives, and false negatives are denoted by green, red, and blue ellipses, respectively.
For each detection, the major and minor axis lengths of the ellipse are set to twice the corresponding FWHM values.
The cyan zoom-in region shows the higher resolution of the reconstructions of POLISH+ and POLISH++ compared to the blurred outputs of CLEAN, underscoring the super-resolution capabilities of our learning-based approach.
Furthermore, the green zoom-in region highlights an area where the original POLISH method introduces artifacts in the background.
In contrast, POLISH++ produces a significantly cleaner, hallucination-free reconstruction.

\begin{figure*}[tb]
\centering
\includegraphics[width=\textwidth]{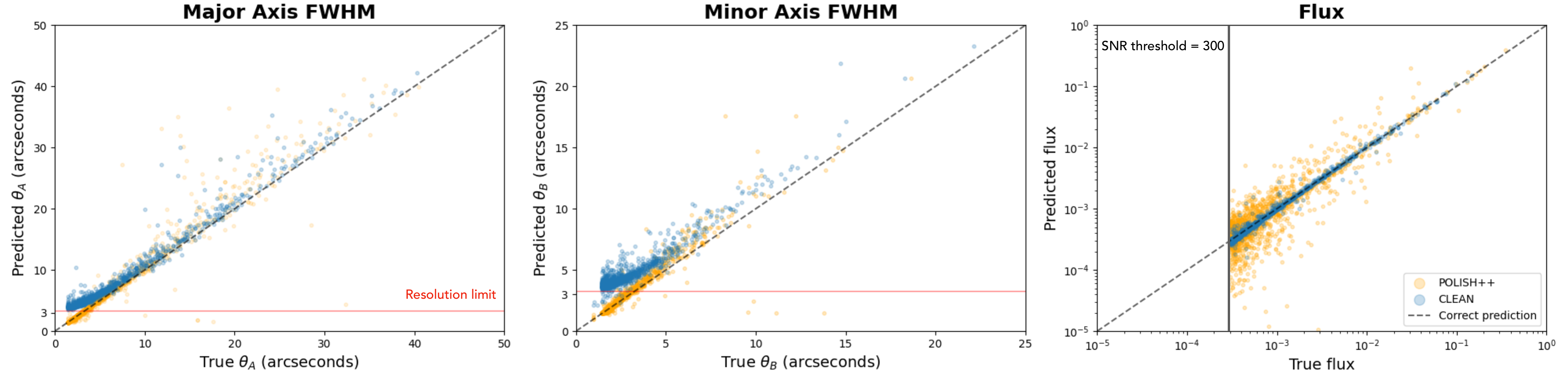}
\caption{
\textbf{Comparison of galaxy property estimation accuracy between CLEAN (blue) and POLISH++~(orange).}
The scatter plots compare the predictions (vertical axes) versus the true values (horizontal axes) for major axis FWHM $\theta_A$ (left), minor axis FWHM $\theta_B$ (middle), and flux (right), across all true positive detections with an SNR threshold of 300.
The dashed line indicates perfect prediction.
POLISH++ produces shape estimates that distribute closer to the diagonal, while CLEAN provides more accurate flux estimates.
It is also clear that CLEAN’s resolving power asymptotes to the PSF angular scale, while POLISH achieves accurate super-resolution below the intrinsic resolution of the PSF ($\approx3.3''$).
}
\label{fig:scatterplot}
\end{figure*}

\noindent \textbf{Source Parameter Estimation} \quad 
We then evaluate the accuracy of estimating the key physical parameters of the detected sources.
Specifically, Tab.~\ref{tab:main} reports the root mean squared errors (RMSE) of the major-axis FWHM ($\theta_A$), minor-axis FWHM ($\theta_B$), and flux, computed from the quantities returned by the source detection algorithm.
To provide a more fine-grained analysis of model behavior across different source properties, Fig.~\ref{fig:scatterplot} presents scatter plots of predicted versus true values of $\theta_A$, $\theta_B$, and flux for CLEAN and POLISH++ on true positive detections with SNR greater than 300.

Our proposed methods substantially improve shape estimation accuracy over the baselines, achieving a significant reduction in RMSE for both $\theta_A$ and $\theta_B$.
In particular, POLISH+ and POLISH++ reduce the major-axis RMSE from $1.0046''$ (CLEAN) to $0.4335''$ and $0.4654''$, respectively, and the minor-axis RMSE from $0.7862''$ (CLEAN) to $0.1889''$ and $0.2056''$.
Consistent with these quantitative improvements, the first two panels of Fig.~\ref{fig:scatterplot} show that the $\theta_A$ and $\theta_B$ estimates produced by POLISH++ distribute much more closely around the unity line than those of CLEAN, indicating significantly improved shape recovery across the full parameter range.

Importantly, for sources with small angular sizes well below the PSF width ($\approx 3.3''$), POLISH++ demonstrates clear super-resolution capability and provides accurate parameter estimates.
In contrast, CLEAN is fundamentally constrained by the PSF and asymptotically approaches the angular size of its central lobe.
This limitation manifests as a flattening trend at small $\theta_A$ and $\theta_B$ values in Fig.~\ref{fig:scatterplot}, and is further illustrated by the visual examples in Fig.~\ref{fig:visual_examples_detections}.

For flux estimation, however, CLEAN remains superior to the learning-based methods, as reflected in both Tab.~\ref{tab:main} and the third panel of Fig.~\ref{fig:scatterplot}.
While POLISH-based methods achieve strong detection and shape recovery performance, their flux RMSE is higher than that of CLEAN.
We attribute this difference to the nonlinear nature of the learning-based reconstruction and the absence of an explicit flux-calibration mechanism.
In contrast, CLEAN operates on the full image in a model-based framework that preserves absolute flux scaling across the full image.
Improving flux fidelity in the learning-based approach—potentially through dedicated calibration steps or post-processing strategies—remains an important direction for future work.

\subsection{Strong Lensing}

\begin{figure}[t]
\centering
\includegraphics[width=\linewidth]{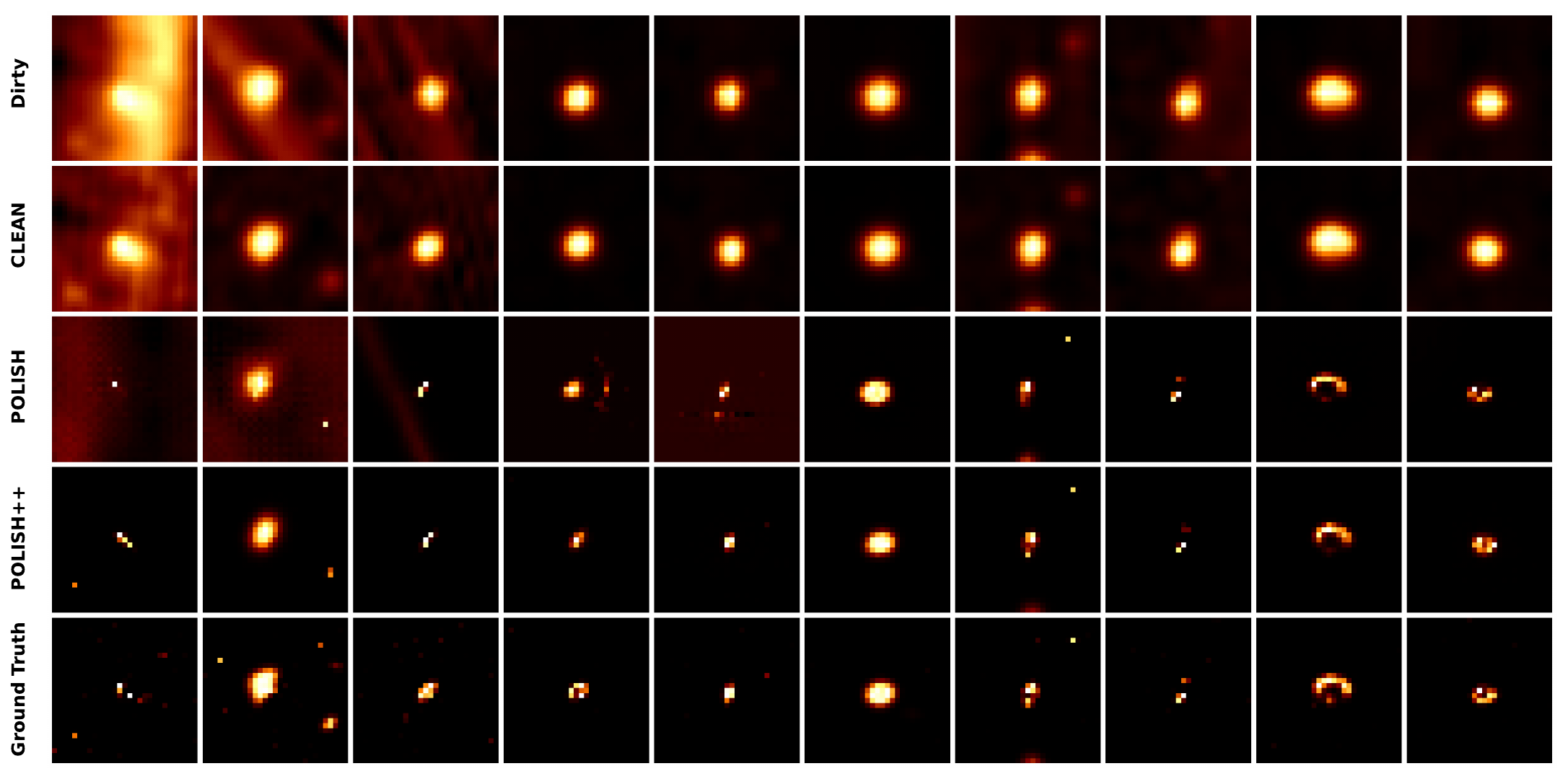}
\caption{Selected example lens cutouts by image type. A lens finding CNN is trained on each image type and used to predict the lens likelihood test images, including the above. The first five lenses were selected to demonstrate the difference between the POLISH and POLISH++ models; the last five were randomly selected. In the first five images, artifacts are apparent in the POLISH row. Both of the POLISH/++ models are clearly able to resolve lensing morphology that is not apparent in the Dirty or CLEAN images.}
\label{fig:strong_lensing_examples}
\end{figure}

Strong gravitational lensing is an increasingly powerful tool for 
probing fundamental physics, cosmology, and astrophysics.
Strong lensing occurs when light from a background source is bent by massive intervening objects, producing multiple copies of the same image. 
While most strong lensing studies have focused on optical and infrared wavelengths, upcoming radio survey telescopes such as the DSA and the SKA will increase the number of radio strong lenses by several orders of magnitude \citep{mckean2015, mccarty2024strong}.
To realize the full potential of strong lensing on the DSA, angular scales at or below the PSF scale must be recovered reliably.
This is because the typical image separation (two times the Einstein radius) of galaxy-galaxy lens systems are 1-2''.
Super-resolution with POLISH offers a path towards a high impact strong lensing survey on the DSA, resulting in as many as $10^5$ new radio lenses \citep{mccarty2024strong}.

A rule of thumb in the literature is that lens discovery requires image separations of $\approx3/2$ times the PSF scale.
But this is at optical wavelengths, where instruments have filled apertures and high fidelity images.
For traditional radio interferometers, poor sampling in the aperture plane (i.e., sparse spatial frequency coverage) leads to stronger requirements on image separation.
\citet{Rezaei2022} find that the maximum angular separation between lensed images needs to be greater than three times the beam FWHM, and the total SNR of the lensed images needs to be greater than 20, for reliable identification by a CNN in simulated International LOFAR Telescope (ILT) data.
We expect that a radio camera like the DSA will have relaxed constraints, comparable to the optical strong lensing case even without super-resolution.
Still, many or most galaxy-scale strong lenses would be likely be missed if a search was done on DSA data using only CLEAN \citep{mccarty2024strong}.
We seek to test if POLISH-enabled super-resolution can aid in the identification of lenses below this limit, thus significantly increasing the strong lens yield of a DSA survey.

To test our method's ability to (super-)resolve strongly lensed radio sources, we first generate a dataset of realistic lensed systems.
We draw sources randomly from the T-RECS simulation \citep{TRECSI,TRECSII}, with a minimum flux-density of $1\,\mu$Jy, maximum apparent angular size of $10''$, and excluding steep-spectrum AGN.
Then, we use the {\tt SL-Hammocks} code\footnote{\url{https://github.com/LSSTDESC/SL-Hammocks} (BSD-3-Clause license)} to randomly draw parameters for the halo-based lens model \citep{Abe_2025}.
This model is desirable because it provides a smooth transition between galaxy, group, and cluster scale systems, which will be important at the image separations we are considering.
The lens redshift, ellipticity, etc. are drawn from calibrated distributions that have been shown to recreate the observed population of strong lenses; we draw the central halo mass from a uniform distribution in log$_{10}(M_{200})$ between 12 and 15 to ensure that we sample large-separation lenses.
We use \texttt{GLAFIC} \citep{Oguri_2010glafic,Oguri_2021} to simulate images of the lens + source system of size $30''\times30''$, after giving the source a random offset in the source plane (but still within the area for multiple imaging in the source plane). 
We then add these objects to the large T-RECS images, assuming random source locations and an areal density that follows a Poisson distribution with mean 30 per deg$^2$, which is the approximate expected number of lenses in a DSA survey, assuming roughly $10^9$ total sources and the CLASS lensing optical depth \citep{CLASSII}.
Dirty, POLISH, POLISH++, and CLEAN data are generated from the T-RECS images as explained above.
When adding the lenses to the T-RECS images, we apply a 10$\times$ multiplier on the lens flux because we wish to isolate the effect of the image separation on lens finding performance without additional complications due to SNR.
This gives a distribution in SNRs that peaks at $\sim$100 and has a minimum of 10. 

\begin{figure}[t]
\centering
\includegraphics[width=0.6\linewidth]{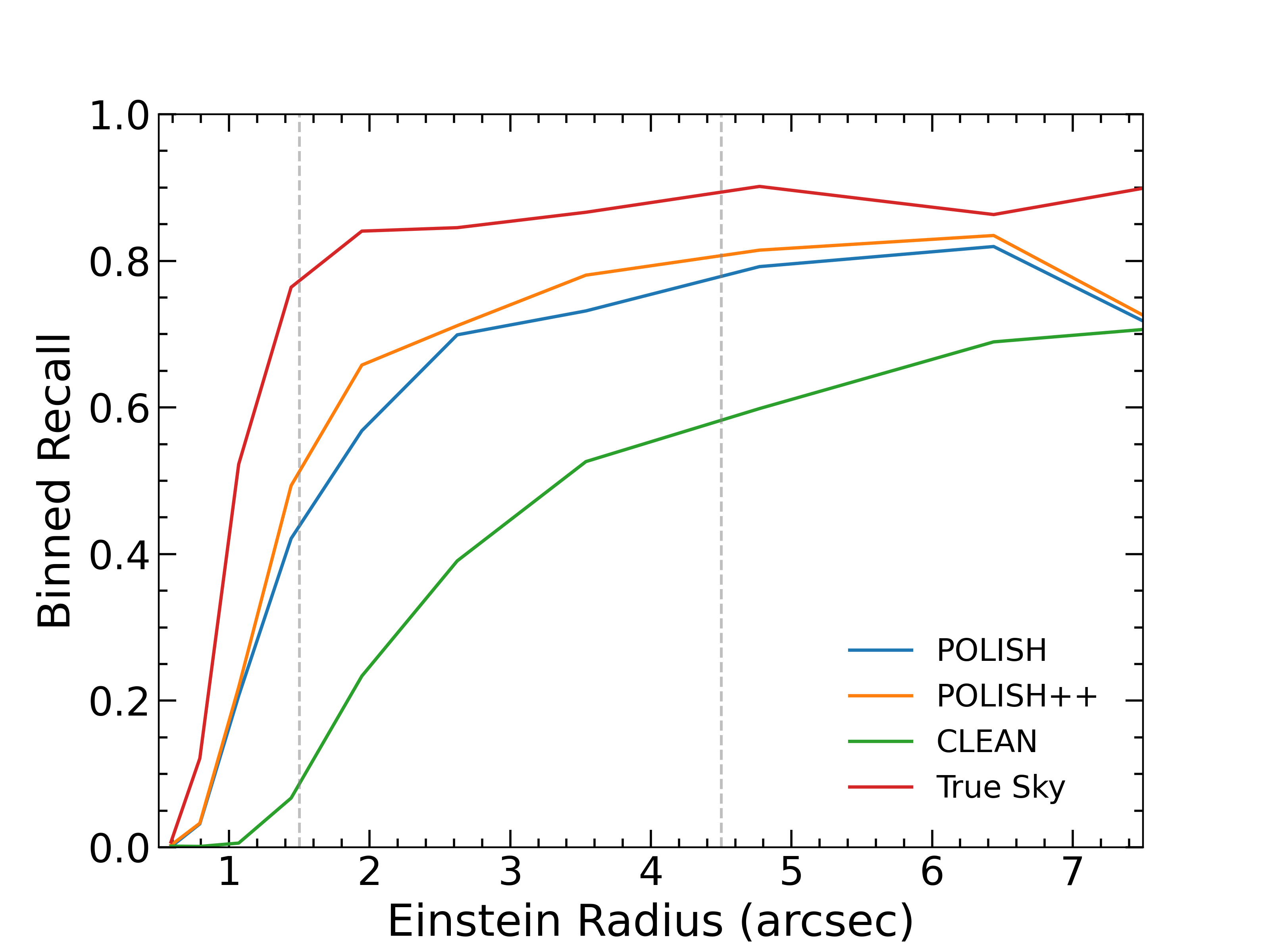}
\caption{
The fraction of lenses that are correctly identified (TPR) by the same lens finder trained/tested on each image type as a function of Einstein radius. The vertical dashed lines correspond to the lens image separation being one and three times the beam FWHM.
}
\label{fig:strong_lensing_line}
\end{figure}

After training the POLISH and POLISH++ models on data with a realistic number of strong lenses, we perform inference on 5 test images of size 12,960$\times$12,960 where the number of strongly lensed sources has been increased by a factor of several, in order to have a large sample of test lenses for the experiment.
We extract $30''\times30''$ cutouts of the lenses, and we use \texttt{PyBDSF} to identify non-lensed sources from these images and extract cutouts of the same size \citep{PyBDSF}.
This gives a total training dataset of 12,702 lensed and 31,429 non-lensed sources.
Next, we train a classifier based on the Structure 1 CNN lens finder from \cite{Rezaei2022} on this dataset, separately for each image type (aka. the ground truth images, dirty images, CLEAN'ed images, POLISH'ed images, and POLISH++'ed images).
We train separate lens finders for each image type because a single lens finder might have trouble generalizing across the image types that have very different resolutions.
In this way we are being conservative, giving each image type the most chance of success. See App.~\ref{app:implementation_details} for more details about the implementation of the lens finder. 
After training, we use the CNNs to predict the lens probability of 8,192 lensed and 28,966 non-lensed cutouts that were not seen previously.
Examples of test lens cutouts are shown in Fig.~\ref{fig:strong_lensing_examples}.
For a chosen false positive rate (FPR) of $10^{-3}$, we show the true positive rate (TPR) for each CNN as a function of the lens Einstein radius (half the maximum image separation) in Fig~\ref{fig:strong_lensing_line}. 

As seen in Fig.~\ref{fig:strong_lensing_examples}, POLISH's super-resolution is able to recover lensing features that are not seen in the dirty or CLEAN images.
The right dashed vertical line in Fig.~\ref{fig:strong_lensing_line} indicates the image separation limit found in \cite{Rezaei2022}, i.e. the lens image separation being 3 times the resolution of the array.
The performance of the CNN trained on standard CLEAN data falls off below this limit, consistent with their findings.
However, the CNNs trained on super-resolved images can recover lenses down to scales close to the PSF FWHM, indicated by the left vertical line.
The expected number of discoverable lenses in a DSA survey increases by roughly an order-of-magnitude from the \cite{Rezaei2022} limit to the left vertical line \citep{mccarty2024strong}.
At all scales, the performance on the POLISH images is remarkably close to the performance on the True Sky images, which we may take as a maximum possible performance for this CNN architecture.
It is worth noting that \cite{Rezaei2022} achieve an overall TPR of $>$90\% for a FPR of less than $10^{-4}$ on their noisy simulated data -- significantly better than our model trained on the ground truth even at large image separations.
This discrepancy is most likely because our cutouts include nearby non-lensed sources.
The density of sources in the DSA images is such that we expect the average cutout to have several unrelated sources detected above 5$\sigma$, some of which may be bright.
These sources will confuse/obscure lenses, or when seen nearby other non-lensed sources could be mistaken as multiple images.
Differences in exact model architecture and training procedures may also contribute to the discrepancy.
Finally, we note that we are fundamentally limited at the very small image separation scales by the $\sim1''$ pixel size of our simulation; even a lens finder trained on the true sky will have difficulty identifying lenses if they are two pixels wide. 

\subsection{Model Mismatch}

We further demonstrate that POLISH++ is both robust and flexible to model mismatches in the PSF $\kbm$, which may arise from practical calibration errors. To evaluate this, we conducted an ablation study investigating how reconstruction performance varies when the training data for the model $G_\theta$ is restricted to images corrupted only by the ideal PSF kernel, $\kbm_{\text{ideal}}$. This analysis allows us to distinguish between performance under perfect model knowledge and performance under realistic model mismatch scenarios. Robustness in the PSF will 
guard against inevitable discrepancies between the 
forward modeled PSF and the true on-sky response. For example, if the ionosphere introduces phase errors that 
are not accounted for, or antennas are slightly mispointed, we want to ensure that POLISH++ can reliably 
deconvolve images. Flexibility will allow us to quickly train models for each new pointing or observing configuration without starting from scratch. 

\begin{figure}[t]
\centering
\includegraphics[width=\linewidth]{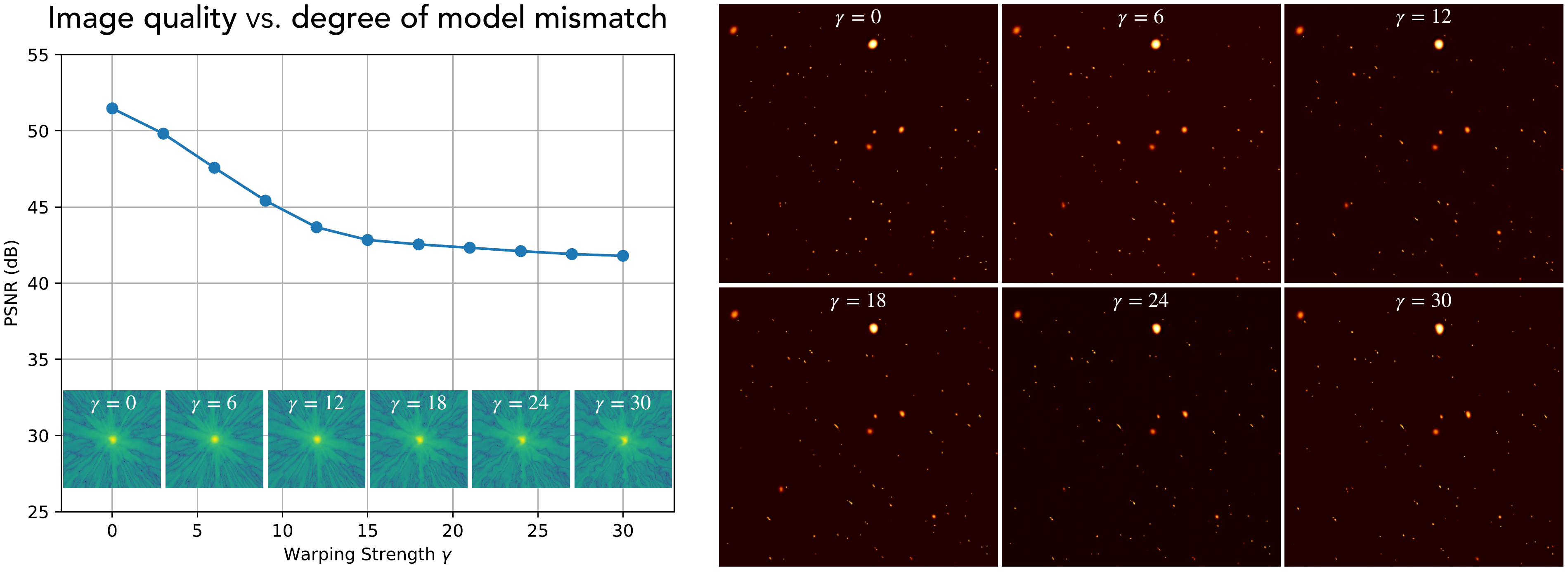}
\caption{
The impact of model mismatch on reconstruction performance. POLISH++ has been trained exclusively on the ideal PSF kernel ($\gamma=0$) and then validated on dirty images corrupted by PSF kernels with increasing warp distortions ($\gamma \in [0, 30]$, moving to the right). The resulting quantitative performance (PSNR) and qualitative reconstructions are displayed in the left and right panels, respectively. 
}
\label{fig:mismatch_robustness}
\end{figure}

\noindent \textbf{Robustness} \quad
In Fig.~\ref{fig:mismatch_robustness}, we evaluate the robustness of POLISH++ by performing image reconstructions on test data where the PSF has been perturbed beyond the distribution encountered during training.
For this experiment, the model was trained exclusively on the ideal PSF kernel ($\gamma=0$), then validated against kernels subject to random warping distortions with strength parameters ranging from $\gamma \in [0, 30]$.
As illustrated in the PSNR performance plot, the quantitative quality of the recovered images predictably degrades as the PSF perturbations increase in severity. 
However, the qualitative performance remains stable; even under extreme distortions ($\gamma=30$), which we do expect in reality, the visual reconstructions stay relatively unchanged.
This discrepancy highlights that while PSNR is highly sensitive to pixel-level shifts and model mismatch, the actual robustness of POLISH++ is better captured by the visual consistency of the recovered sources.

\begin{figure}[t]
\centering
\includegraphics[width=0.6\linewidth]{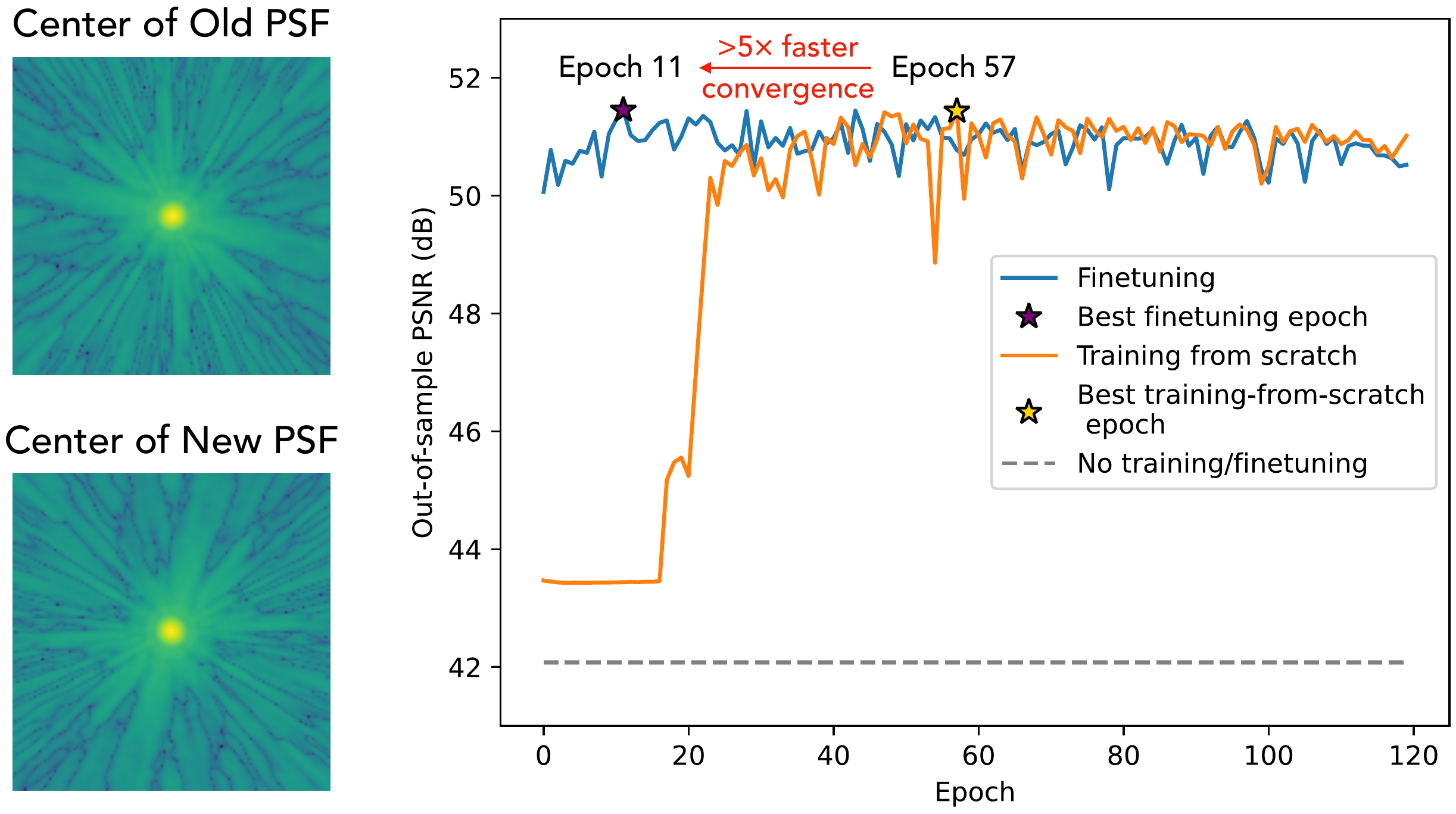}
\caption{
Adaptation flexibility and convergence acceleration of POLISH++ under model mismatch. While both fine-tuning and training from scratch significantly outperform the baseline model without adaptation (dashed gray line), fine-tuning an existing model (top left) to a new PSF distribution (bottom left) yields a substantial increase in training efficiency.
}
\label{fig:mismatch_flexibility}
\end{figure}

\noindent \textbf{Flexibility} \quad
Beyond static robustness, we evaluate the flexibility of POLISH++ in adapting to entirely new system parameters. 
In Fig.~\ref{fig:mismatch_flexibility}, we compare the convergence speeds of fine-tuning an existing model versus training a new one from scratch when encountering a mismatched PSF.
While both approaches eventually yield significantly higher PSNR values than directly applying the mismatched original model, fine-tuning provides a substantial acceleration in convergence.
Notably, the fine-tuning process reaches its peak performance in 11 epochs, whereas training from scratch requires 57 epochs to achieve its best result.
This represents a more than five-fold speedup, demonstrating that POLISH++ enables rapid adaptation to diverse instrument configurations by leveraging previously learned representations. 
As a real-world example, an interferometric array's PSF varies as a function of sky position, even though the properties of that PSF vary slowly and smoothly with pointing.
We can train a large baseline model and finetune it for different pointings, observing conditions, and per-antenna performance. 
\section{Conclusion}

We have presented a set of practical advances that bring deep learning-based radio interferometric imaging closer to real deployment on next-generation survey instruments such as the DSA.
By introducing patch-wise training and inference to handle extremely large images, and an inverse-hyperbolic-function-based transform to handle the high dynamic range of realistic skies, we introduce POLISH++ to scale far beyond previous deep-learning approaches. 
Using large, realistic T-RECS images, we demonstrated improved reconstruction fidelity and source detection accuracy, as well as robustness and adaptivity to PSF calibration errors.
We further showed that the proposed method can super-resolve strongly lensed systems, a capability that could significantly increase the strong-lens yield of future wide-field surveys.
Together, these results indicate that properly designed learning-based methods can meet key challenges in real-world radio interferometric imaging and play a promising role in future analysis pipelines.

%% Also note that the akcnowlodgment environment does not support long amounts of text. If you have a lot of people and institutions to acknowledge, do not use this command. Instead, create a new \section{Acknowledgments}.
\begin{acknowledgments}
We thank Viviana Rosero, Joshua Albert, and Casey Law 
for providing forward-modeled point spread functions for the Deep Synoptic Array 
and for valuable discussions on the measurement process.
\end{acknowledgments}


% \vspace{5mm}
% \facilities{HST(STIS), Swift(XRT and UVOT), AAVSO, CTIO:1.3m,


\software{astropy \citep{2013A&A...558A..33A,2018AJ....156..123A},  
          Source Extractor \citep{1996A&AS..117..393B}
          }



\appendix
\section{Detailed Explanations for Tab.~\ref{TAB:COMPARISON}} \label{app:table_explanation}

Here we provide more details for Tab.~\ref{TAB:COMPARISON}.

For POLISH \citep{connor2022deep}, the maximum image size $2,048$ is given in Sec. 4.1 of the paper and the open-sourced \href{https://github.com/liamconnor/polish-pub}{repository}.
To get the maximum dynamic range, we first generate a dataset using the given commands.
We find that the maximum pixel intensity before normalization is around $2,000$.
Noting that the standard deviation of the background noise is $5$ as shown \href{https://github.com/liamconnor/polish-pub/blob/main/make_img_pairs.py#L113}{here}.
Assuming that the weakest detectable galaxy is $\sim4$ standard deviations, the dynamic range is around $\frac{2000}{4*5}=100$.
The paper considered weak lensing and generalization under PSF mismatch due to synthetic warping.

For Radionets, the maximum image sizes are given in Tab. 4 of \citep{schmidt2022deep} ($1,024$) and Sec. 2 of \citep{geyer2023deep} ($128$), respectively.
To get the maximum dynamic range, we inspect the \href{https://github.com/radionets-project/radionets/blob/main/radionets/simulations/utils.py#L206-L226}{data simulation code} and find that the standard deviation of background noise is $0.05\cdot m$ where $m$ the maximum of the clean sky image.
Assuming that the weakest detectable galaxy is $\sim4$ standard deviations, the dynamic range is around $\frac{m}{4*0.05m}=5$.

For Deflation Net \citep{chiche2023deep}, the maximum image size $256$ and maximum dynamic range $100$ are both given in Sec. 5.1.2 of the paper.

For R2D2 \cite{aghabiglou2024r2d2}, the maximum image size $512$ and maximum dynamic range $5\cdot10^5$ are given in Sec. 3.2 and Sec. 3.3 of the paper, respectively.

For GU-Net \cite{mars2025learned}, the maximum image size $512$ and maximum dynamic range $\sim600$ are given in Sec. 3.4 and Sec. 4.3 of the paper, respectively.

For RI-GAN \cite{mars2025generative}, the maximum image size $360$ and maximum dynamic range $\sim600$ are given in Sec. 4.3 and Sec. 5.3 of the paper, respectively.

\section{Implementation Details}
\label{app:implementation_details}

\noindent \textbf{Forward model} \quad
Throughout this work, we consider the forward model (\ref{eqn:forward_model}) with an RMS noise $\nbm \sim \Ncal(\bm{0}, \sigma^2\Ibm)$ with $\sigma=10^{-6}$ (i.e., 1 $\mu$Jy) and a downsampling factor of $s=2$.
We use a single DSA-1650 PSF $\kbm$ described in Sec.~\ref{sec:experimental_setup} for generating the training data.
Note that this is different from \cite{connor2022deep} where a distribution of PSFs was used for training.

\noindent \textbf{Model training} \quad
We trained all POLISH, POLISH+, POLISH++ models in this work using the Adam optimizer \citep{kingma2014adam} with a learning rate of $0.0001$ and a batch size of $12$ with the loss function defined in (\ref{eqn:loss}).
We set $a_\text{dirty}=a_\text{true}=0.1$ for all the experiments.

\noindent \textbf{Detection matching} \quad
When matching detected galaxies to ground-truth objects, we set a galaxy-specific detection threshold based on the size of the closest ground-truth detection to which it is matched.
Let $\theta_A$ and $\theta_B$ denote the semi-major and semi-minor axes of the ground-truth galaxy, respectively.
We define the matching threshold as
$$N_\text{threshold} = 2\sqrt{\theta_A \theta_B}.$$
The rationale for this choice can be understood by considering the special case of circular galaxies with equal size, for which $\theta_A = \theta_B = r$.
In this case, two such galaxies have nonzero overlap only when the distance between their centers is less than $2r$, which coincides with $N_\text{threshold} = 2\sqrt{\theta_A \theta_B}$.
This definition therefore provides a size-adaptive matching criterion.

\noindent \textbf{Lens finder} \quad The CNN lens finders are based on the architecture of the Structure 1 in \cite{Rezaei2022}. This architecture consists of three stages. First, a multi-scale filter bank which convolves the input image with kernels of size 1x1, 3x3, and 5x5 in parallel and concatenates the output. Second, an inception block where the data takes three paths, a further 1x1 convolution, 1x1 $\rightarrow$ 3x3, or 1x1 $\rightarrow$ 3x3 $\rightarrow$ 5x5. The purpose of these steps with kernels of different sizes is to capture lensing features that can appear on multiple scales. Finally, the output feeds into two dense fully connected nerural network layers, which end in a single binary classification node with output between 0 and 1. The total number of model parameters for each CNN is 28,188,921.
During training, we use rotations and flips to augment the data. We use biased sampling to ensure that the CNNs see a balanced number of lensed and non-lensed sources in each training epoch, because there are a larger number of non-lensed sources in our training set.
Because the training set is also heavily biased towards small separation lenses, we also find it necessary to oversample the large separation lenses during training. We use the Binary Cross Entropy (BCE) loss function and the Adam optimizer \citep{kingma2014adam}. The learning rate is initially set as 0.001, but we reduce the learning rate during training in the validation loss plateaus. 

\noindent \textbf{Visualization} \quad
Due to the high dynamic range of the T-RECS images and the DSA PSF, we applied the following transformations to the figures with reconstructions and PSFs for better visual quality:
\begin{itemize}
    \item Fig.~\ref{fig:patch_lr}:
        \begin{itemize}
            \item Full-FOV measurement ($12,960\times12,960$ pixels): $\mathsf{AsinhStretch}(\,\cdot\,; 10^{-8})$
            \item Ground truth patch ($1,296\times1,296$ pixels): $\mathsf{AsinhStretch}(\,\cdot\,; 10^{-7})$
            \item Patch measurements and CLEAN patch ($1,296\times1,296$ pixels): $\mathsf{AsinhStretch}(\,\cdot\,; 10^{-6})$
        \end{itemize}
    \item Fig.~\ref{fig:visual_examples_detections} (all images have $324\times324$ pixels):
        \begin{itemize}
            \item Ground truth, POLISH, and POLISH++ reconstructions: $\mathsf{AsinhStretch}(\,\cdot\,; 10^{-7})$
            \item Noisy ground truth, measurement, and CLEAN reconstruction: $\mathsf{AsinhStretch}(\,\cdot\,; 10^{-5})$
        \end{itemize}
    \item Fig.~\ref{fig:strong_lensing_examples} (all images have $39\times39$ pixels): min-max normalization. 
    \item Fig.~\ref{fig:mismatch_robustness}:
        \begin{itemize}
            \item PSFs ($128\times128$ pixels): $\ln(\operatorname{abs}(\,\cdot\,))$
            \item Reconstructions ($324\times324$ pixels): $\mathsf{AsinhStretch}(\,\cdot\,; 10^{-3})$
        \end{itemize}
    \item Fig.~\ref{fig:mismatch_flexibility}:
        \begin{itemize}
            \item PSFs ($128\times128$ pixels): $\ln(\operatorname{abs}(\,\cdot\,))$
        \end{itemize}
\end{itemize}
Moreover, for Fig.~\ref{fig:asinh}, the dynamic range in panel (b) is computed by dividing the brightest pixel in each patch by $4\times10^{-6}$ (i.e., 4$\mu$Jy).
We adopt this value because $4\times10^{-6}$ corresponds to the $4\sigma$ detection threshold and is therefore representative of the lowest intensity at which galaxies are considered detectable.
The actual minimum value in each patch is highly close to 0, due to the existence of noise and the use of float32 precision, so defining the dynamic range using the actual minimum pixel value would result in artificially inflated and misleading values.

\bibliography{references}{}
\bibliographystyle{aasjournal}



\end{document}